\documentclass[12pt]{iopart}
\usepackage{iopams}  
\begin{document}

\title[Construction of Hamiltonian and Nambu Forms for the Shallow Water Equations
]{Construction of Hamiltonian and Nambu Forms for the Shallow Water Equations
}

\author{Richard Blender}
\address{Meteorological Institute, University of Hamburg, 
Hamburg, Germany}
\ead{richard.blender@uni-hamburg.de}
\author{Gualtiero Badin}
\address{Institute of Oceanography, Center for Earth System Research and Sustainability (CEN), University of Hamburg, Hamburg, Germany}
\ead{gualtiero.badin@uni-hamburg.de}	

\begin{abstract}
A systematic method to derive the Hamiltonian and Nambu form for the shallow water equations, using the conservation for energy and potential enstrophy, is presented.
Different mechanisms, such as vortical flows and emission of gravity waves, emerge from different conservation laws (CLs) for total energy and potential enstrophy. The equations are constructed using exterior differential forms and self-adjoint operators and result in the sum of two Nambu brackets, one for the vortical flow and one for the wave-mean flow interaction, and a Poisson bracket representing the interaction between divergence and geostrophic imbalance. 
The advantage of this approach is that the Hamiltonian and Nambu forms can be here written in a coordinate independent form.

\end{abstract}

%version 12

\today

\maketitle

% now start line numbers
%\linenumbers

\section{Introduction}

Noncanonical Hamiltonian dynamics, which is the natural framework for the geometric description of hydrodynamical systems in Eulerian form, is characterized by the fact that the Poisson operator is singular, with the singularity giving rise to a class of conserved quantities called Casimirs. Following the existence of these conserved quantities, it is possible to reformulate the dynamics from the classical formulation that instead makes use only of the Hamiltonian.  
The resulting dynamics based on this assumption is simply based on Liouville's Theorem and was first developed by  Nambu \cite{Nambu1973}. Differently from Hamiltonian dynamics, Nambu dynamics can be odd-dimensional  and it results in a skew-symmetric bracket, now called Nambu bracket, which obeys discrete Leibniz Rule and the Jacobi Identity \cite{Takhtajan1994}.
The approach has been applied to finite-dimensional systems, including the nondissipative Lorenz equations \cite{NevirBlender1994} and the dynamics of point-vortices \cite{Makhaldiani07,Makhaldiani12,MuellerNevir14}, and it has been extended to infinite dimensional systems by N\'{e}vir and Blender \cite{NevirBlender1993}, where the Nambu brackets for ideal hydrodynamics were formulated using enstrophy and helicity as conserved quantities in two and three dimensions respectively. Infinite dimensional Nambu dynamics has been applied to geophysical fluid dynamics \cite{Bihlo2008,GassmannHerzog2008,NevirSommer2009,SommerNevir2009,SalazarKurgansky2010,BlenderLucarini2013}, where the resulting formulation can be used not only to provide an invariant framework, but also for the construction of numerical algorithms that conserve both the energy and the Casimirs of the system \cite{Salmon2005, bridges2006numerical, Salmon2007,GassmannHerzog2008,sommer2010phase}. The downside of the construction of the hydrodynamics Nambu-brackets is that, as much as for the construction of the hydrodynamics Poisson brackets \cite{MorrisonGreene80}, it was mainly based on intuition and guessing.

Recently, \cite{BlenderBadin15} showed however that the equations for hydrodynamic systems can be derived in Nambu form purely using conservation laws (CLs) and geometric principles. In this derivation, the CLs are considered as constraints for the trajectories in state space. For two-dimensional ideal hydrodynamics with vorticity as the single degree of freedom and the conservation integrals of kinetic energy and enstrophy, this approach yields the vorticity equation. If a temperature perturbation is introduced as a second degree of freedom and if enstrophy is replaced by a conserved integral for the product of vorticity and this new second variable, one obtains the Rayleigh-B{\'e}nard equations. The dynamics can be described in terms of physical processes if the second integral is replaced by two integrals making use of the squares of vorticity and temperature, which correspond to enstrophy and available potential energy. This separation yields modular equations. Such integrals, which have been called constitutive laws \cite{NevirSommer2009,SalazarKurgansky2010}, determine the dynamics and are not conserved in the full system.

Here we apply a similar approach for the derivation of the shallow water equations. While the Nambu form of these equations was obtained by Salmon \cite{Salmon2007}, the equations are here re-derived through the application of geometrical methods rather than from intuition. In Section \ref{sec:1} we present the shallow water equations and the associated conserved quantities. In Section \ref{sec:2} we construct the different dynamical equations through different approximations of the conservation of energy and of the Casimirs of the system. In Section \ref{sec:3} we present the Nambu brackets for the shallow water equations. Finally, in the Appendix, we present the Nambu form for a finite-dimensional system obtained by the coefficients of the Fourier expansion of the shallow water equations and describing the interaction between the "slow" vorticity variables and the "fast" gravity waves modes.

%%%%%%%%%%%%%%%%%%%%%%%%%%%%%%%%%%%%%%%%%%

%The method yields geometric trajectories in state space and the CLs are exactly represented by the equations. 
%However, the temporal evolution along the trajectory cannot be derived and the construction is not unique.

%====================================
\section{The Shallow Water Equations}
\label{sec:1}
%====================================

Consider a fluid in a single layer with constant density and under the effect of the Earth's rotation. The fluid is between a flat bottom boundary and a free surface at $z = h(x,y)$. We set the density to be uniform, i.e. $\rho=1$. The dynamics of the horizontal flow $\vec{v}= (u, v)$ can be written in terms of the vorticity 
\begin{equation}
\zeta = \hat{k} \cdot \nabla \times \vec{v} = \frac{\partial v }{ \partial x} - \frac{\partial u }{ \partial y}~, 
\label{eq:N1}
\end{equation}
divergence 
\begin{equation}
\mu = \nabla \cdot \vec{v} = \frac{ \partial u }{ \partial x} + \frac{ \partial v }{ \partial y}~,
\label{eq:N2}
\end{equation} 
 and height $h$. Following \cite{Salmon2007}, we write the momentum in terms of a stream-function $\chi$, and a potential $\gamma$,  so that 
 \begin{eqnarray}
 h \vec{v} & = \left(- \frac{ \partial \chi }{ \partial y} + \frac{ \partial \gamma }{ \partial x } ~,~ \frac{ \partial \chi }{ \partial x} + \frac{ \partial \gamma }{ \partial y} \right)  \nonumber \\
 &= \vec{k} \times  \nabla \chi  + \nabla \gamma~. 
 \label{eq:N3}
\end{eqnarray} 
 The resulting shallow water potential vorticity is  
  \begin{eqnarray}
 q = \frac{\zeta +f}{h}~, 
  \label{eq:N4}
\end{eqnarray}
 where $f$ is the Coriolis frequency.

The system conserves two integrals, namely the total energy
\begin{equation}
H = \frac{1}{2} \int [ h v^2 + g h^2 ] dA~,
\label{eq:2.1}
\end{equation}
and the 
%potential enstrophies \cite{Salmon2005}
%\begin{equation}
%Z_n =  \frac{1}{2}  \int_A  h q^{2+n} dA~. 
%\label{eq:N5}
%\end{equation}
%In this study we will consider only the $n=0$ 
potential enstrophy 
\begin{equation}
Z =  \frac{1}{2} \int  h q^2 dA~.
\label{eq:2.2}
\end{equation}
Notice that the system conserves also the n-order potential enstrophies \cite{Salmon2005}
\begin{equation}
Z_n =  \frac{1}{2+n}  \int_A  h q^{2+n} dA~, 
\label{eq:N5}
\end{equation}
where only the $n=0$ case will be considered in this study.
Finally, the system conserves also mass.

The functional derivatives of $H$ and $Z$ with respect to the dynamical variable are abbreviated by subscripts, e.g. $\delta H / \delta \zeta = H_{\zeta}$, and yield 
\begin{eqnarray}
& H_{\zeta}  = - \chi~,   \label{eq:2.3a} \\
& H_{\mu} = - \gamma~,   \label{eq:2.3b} \\
& H_h =  \Phi  ~,  \label{eq:2.3c} 
\end{eqnarray}
where
\begin{equation}   
\Phi  = \frac{1}{2} v^2 + g h~,
\label{eq:2.4}
\end{equation}
is a generalized Bernoulli function and
\begin{eqnarray}
& Z_{\zeta} = q~,   \label{eq:2.5a} \\
& Z_{\mu} = 0~,   \label{eq:2.5b} \\
& Z_h = - \frac{1}{2} q^2   ~. \label{eq:2.5c}
\end{eqnarray}
Notice that the functional derivatives can be easily solved making use of (\ref{eq:N3}).

%====================================
\section{Construction of the Dynamical Equations}
\label{sec:2}
%====================================

To construct the dynamical equations, we will use the conservation of the total energy, 
\begin{equation}
\frac{dH}{dt}	=  \int \left[ H_{\zeta} \frac{\partial \zeta}{\partial t} + H_{\mu} \frac{\partial \mu}{\partial t} + H_h  \frac{\partial h}{\partial t}  \right] dA  = 0~,
\label{eq:3.1}
\end{equation}
and potential enstrophy 
\begin{equation}
\frac{dZ}{dt}	=  \int \left[ Z_{\zeta} \frac{\partial \zeta}{\partial t} + Z_{h} \frac{\partial h}{\partial t}  \right] dA  = 0~.
\label{eq:3.2}
\end{equation}

Further, we will use the self-adjoint operator $\mathcal{A}$ defined as
\begin{equation}
\mathcal{A}f = \nabla \cdot (w \nabla f)~,
\label{eq:3.15}
\end{equation}
where $w$ is a yet undetermined function. The symmetry of the operator $\mathcal{A}$ is used to determine the energy balance
 \begin{equation}
\int \left[ H_j \mathcal{A} H_i  - H_i \mathcal{A} H_j  \right] dA = 0~,
\label{eq:3.16}
\end{equation}
where, once again, subscripts $i,j \in \{ h, \mu, \zeta \}$ indicate functional derivatives. 

%The maintenance of these CLs is the basis for the derivation of the dynamical equations. 
In the following, we will use individual terms of the CLs  and different choices for the self adjoint operator to construct the dynamical equations for physical processes. Throughout we will assume periodic boundary conditions. 
%The identification of the dynamic equations is not unique. If we consider both equations as a linear system of equations with rank two, we have one free parameter. 
%The equations are energetically consistent. This approach leads directly to Hamiltonian and Nambu formulations. We construct terms for individual processes by the restriction to relevant energy exchange. For the full equations, all terms are added, but for process studies they can be analysed separately. The terms can be found as Eqs. (2.8) in \cite{Salmon2007}.

%====================================
\subsection{Vorticity Advection}
%====================================

The vorticity equation can be derived straightforwardly from the previous definitions. 
Assuming that the flow is purely vortical, 
we neglect the evolutions of the divergence and the height
%the assumptions of nondivergence $\mu = 0$ 
%  and the rigid lid upper boundary condition $\partial h / \partial t = 0$ yield, from 
in (\ref{eq:3.1}) and (\ref{eq:3.2})
%
%The advection term is based on the velocity, comprised of the sum of the non-divergent vortical part of the flow, which is described by the momentum stream-function $\chi$, and the divergent part the flow, which is described by the velocity potential $\gamma$. The advected vorticity is given by the functional derivative of potential enstrophy, which appears in (\ref{eq:2.5a}). 
%%These are terms using $H$ and $Z$ functional derivatives (mixed terms). 
%
%Using only the first two terms of (\ref{eq:3.1}) and (\ref{eq:3.2}) yields
\begin{equation}
\frac{dH}{dt}	=  \int H_{\zeta} \frac{\partial \zeta}{\partial t}  dA  = 0~,
\label{eq:3.3}
\end{equation}
\begin{equation}
\frac{dZ}{dt}	=  \int Z_{\zeta} \frac{\partial \zeta}{\partial t} dA  = 0~. 
\label{eq:3.4}
\end{equation}
Following \cite{BlenderBadin15}, 
consider the exact 2-form $df \wedge dg$, where $d$ is the exterior differential operator, $\wedge$ is the wedge product, $f$ and $g$ are arbitrary functions, $f,g : \mathbb{R}^2 \rightarrow \mathbb{R}$ that depend on the phase space variables, 
and $dA = dx \wedge dy $. 
To construct the dynamics we thus use $\int df \wedge dg =0$, so that
\begin{equation}
\int f ~df \wedge dg = 0, ~  \int g ~ df \wedge dg = 0  ~.
\label{eq:3.5}
\end{equation}
In our application $f = H_{\zeta}$, $g = Z_{\zeta}$, hence, from (\ref{eq:3.3}), (\ref{eq:3.4}) and (\ref{eq:3.5}),
%Using these equations we can eliminate the first terms in (\ref{eq:3.1})-(\ref{eq:3.2}). This yields the vorticity equation 
\begin{equation}
\frac{\partial \zeta}{\partial t} dx \wedge dy = dH_{\zeta} \wedge dZ_{\zeta} = J(q, \chi) dx \wedge dy~,
\label{eq:3.6}
\end{equation}
where $J(a,b) = \partial a / \partial x ~\partial b / \partial y ~-~  \partial a / \partial y~ \partial b / \partial x$ is the Jacobian operator,  and where (\ref{eq:2.3a}) has been used. It should be noted that (\ref{eq:3.6}) is determined up to a constant factor. In this way, the method here introduced is set for a \emph{construction} rather than a \emph{derivation} of the shallow water equations. The analogy is explained by the definition of the CLs as vertical integrals with a height $h$. Notice that this derivation of the vorticity equation was independent on the use of the self-adjoint operator $\mathcal{A}$.

The advection of vorticity has a Nambu representation with enstrophy as a second conserved quantity used for the derivation of the dynamics \cite{NevirBlender1993}. The Nambu representation shows that this equation can be considered as the most elementary hydrodynamic system with two CLs \cite{BlenderBadin15} and will be re-derived in Section \ref{sec:3}.
%
%% %====================================
%% \subsection{Divergent Component of the Flow}
%% %====================================
%% (??????) If we assume $\partial \mu / \partial t =0$ and a rigid lid 
%% $\partial h / \partial t = 0$ both CLs can be satisfied.
%% In this case, the second term of (\ref{eq:3.1}) can be set to zero in combination with $Z_{\zeta}$, so that
%

To determine the interaction between the vortical and the divergent components of the flow, we are restricted to a single
%re-insert the 
term depending on $\mu$ in (\ref{eq:3.1}).
The enstrophy conservation does not constrain the divergence evolution 
% and (\ref{eq:3.4}), obtaining
% \begin{eqnarray}
\begin{equation}
\frac{dH}{dt}	=  \int %% \left[ H_{\zeta} \frac{\partial \zeta}{\partial t} + 
H_{\mu} \frac{\partial \mu}{\partial t}  %% \right] 
dA  = 0~.  \label{eq:3.7a} \\
\end{equation}
This is satisfied by
\begin{equation}
\frac{\partial \mu}{\partial t} dx \wedge dy = dH_{\mu} \wedge dZ_{\zeta} = J(q, \gamma) dx \wedge dy~,
\label{eq:3.8}
\end{equation}
where we used $Z_{\zeta}$ as a choice for the arbitrary function satisfying (\ref{eq:3.7a}) in order to ensure the interaction between $\gamma$ and $q$.
Furthermore, this shows that the functional derivatives of the Hamiltonian, $H_{\zeta}$ and $H_{\mu}$, are responsible for the hydrodynamic transport, 
i.e. the advection, while the functional derivative of $Z$ represents the advected quantity. 
% \frac{dZ}{dt}	=  \int Z_{\zeta} \frac{\partial \zeta}{\partial t} dA  = 0~. 
% \label{eq:3.7b}
% \end{eqnarray}
% which is satisfied by
%% \begin{equation}
%% \frac{\partial \mu}{\partial t} dx \wedge dy = dH_{\mu} \wedge dZ_{\zeta} 
%% =  J(q, \gamma) dx \wedge dy~,
%%\label{eq:3.8}
%% \end{equation}
%% This  describes the transport of vorticity with the divergent momentum 
%% potential flow by the potential $\gamma$. 
%% From the relations (\ref{eq:3.6}) and (\ref{eq:3.8}) one sees that 
%% us, both terms are mixed in the sense that they use both, 
%% $H$ and $Z$ derivatives.
%% the two functional derivatives of the Hamiltonian, 
%% $H_{\zeta}$ and $H_{\mu}$, 
%% are responsible for the hydrodynamic transport, 
%% i.e. the advection, while the functional derivative of 
%% $Z$ represents the advected quantity. 

%====================================
\subsection{Gravity Waves}
%====================================
In order to derive the dynamical equations associated with the gravity waves, we start by neglecting the vortical flow represented by the terms $H_{\zeta}$ and $Z_{\zeta}$. 
% by setting $\chi=0$ and $q=0$, that means $H_{\zeta}=0$ and $Z_{\zeta}=0$ respectively. 
Under these assumptions (\ref{eq:3.1}) yields
\begin{equation}
\frac{dH}{dt}	=  \int \left[ H_{\mu} \frac{\partial \mu}{\partial t} + H_h \frac{\partial h}{\partial t} \right] dA  =0 ~.
\label{eq:3.9}
\end{equation}
To construct the dynamical equations for $\mu$ and $h$ we identify the self-adjoint operator with the Laplacian  operator
\begin{equation}
\mathcal{A} (.) = \nabla^2 (.)~. 
\label{eq:N6}
\end{equation}
Observing that the energy is conserved for
 \begin{equation}
\int \left[ H_{\mu} \nabla^2 H_h - H_h \nabla^2 H_{\mu} \right] dA = 0~,
\label{eq:3.10}
\end{equation}
we apply this symmetry property to cancel $H_h \partial h / \partial t$  and  $H_{\mu} \partial \mu / \partial t$ in the energy conservation (\ref{eq:3.9}). 
%The dynamical equations are defined only up to a constant factor. 
Identifying the Laplacian terms as time derivatives we obtain 
\begin{eqnarray}
\frac{\partial \mu}{\partial t} = - \nabla^2 H_h = - \nabla^2 \Phi~,  \label{eq:3.11} \\
\frac{\partial h}{\partial t}  =  \nabla^2 H_{\mu} =  - \nabla^2 \gamma~. \label{eq:3.12}
\end{eqnarray}
The resulting energy exchange is related to the divergence $\mu$ and thus to gravity waves. Notice that the choice of the Laplacian was an \emph{ad hoc} choice. Also for this, the method here introduced is set for a construction of the shallow water equations. 
%Note that these terms can be added in the shallow water equations, but could be considered individually. 

%Note that it is not possible to satisfy both CLs with a single process. For example, when $\partial h / \partial t$ does not vanish for gravity waves, $Z$ cannot be constant, without vorticity changes, $\partial \zeta / \partial t$, (since $q$ is not zero), therefore we need further processes for $dZ/dt$.

%
%====================================
\subsection{Vortical Component of the Flow and Gravity Waves Interaction}
%====================================
%
The interaction between vortical and divergent flows couples the quasi-two-dimensional large-scale flow and the gravity wave field in geophysical fluid dynamics. 
%% This is the basic mechanism for the emission of gravity waves by 
%% unbalanced flows or by the equilibration of a balanced flow \cite{Vanneste13}. 
To study this interaction, we impose that the interaction between the 
different components of the flow is \emph{not} responsible 
for changes in the \emph{total} (kinetic plus potential) energy of the flow. 
%% Setting thus $H_h = \Phi = 0$, (\ref{eq:3.1}) and (\ref{eq:3.2}) yield
Hence we neglect $H_h = \Phi$ in (\ref{eq:3.1}) %% and (\ref{eq:3.2}) yield
\begin{equation}
\frac{dH}{dt}	
=  \int \left[ H_{\zeta} \frac{\partial \zeta}{\partial t} 
+ H_{\mu} \frac{\partial \mu}{\partial t}   \right] dA  = 0~, \label{eq:3.13a} 
%% \label{eq:3.14}
\end{equation}
%
%To construct the dynamical equations we begin with the equation for $H$ and use the self-adjoint operator 
%\begin{equation}
%\mathcal{A}f = \nabla \cdot (w \nabla f)~,
%\label{eq:3.15}
%\end{equation}
%where $w$ is a yet undetermined function. The symmetry of the operator $\mathcal{A}$ is used to determine the energy balance
% \begin{equation}
%\int \left[ H_{\zeta} \mathcal{A} H_{\mu}  - H_{\mu} \mathcal{A} H_{\zeta}  \right] dA = 0~.
%\label{eq:3.16}
%\end{equation}
From (\ref{eq:3.13a}) %%-(\ref{eq:3.13b}) 
and (\ref{eq:3.16}) we identify the evolution terms as
\begin{eqnarray}
\frac{\partial \zeta}{\partial t} = \mathcal{A} H_{\mu} ~, \label{eq:3.17} \\
\frac{\partial \mu}{\partial t}  =  - \mathcal{A}  H_{\zeta} ~.  \label{eq:3.18}
\end{eqnarray}
The function $w$ in (\ref{eq:3.15}) can now be determined by the conservation of 
potential enstrophy (\ref{eq:3.2}) with the functional derivatives  (\ref{eq:2.3c})-(\ref{eq:2.5a}) 
\begin{equation}
\frac{dZ}{dt}	=  \int \left[ Z_{\zeta} \frac{\partial \zeta}{\partial t} + Z_{h} \frac{\partial h}{\partial t}  \right] dA  = 0~. % \label{eq:3.13b}
\label{eq:3.14}
\end{equation}
For the $h$-dynamics we use  (\ref{eq:3.12}),  obtained for gravity waves above. 
With the vorticity dynamics (\ref{eq:3.17}) we find
\begin{eqnarray}
\frac{dZ}{dt}	=  \int \left[ q \nabla \cdot \left( w \nabla H_{\mu} \right) - \frac{1}{2} q^2 \nabla^2 H_{\mu}  \right] dA   \nonumber \\
= \int \left[ q \left( \nabla w \cdot \nabla H_{\mu} + w \nabla^2 H_{\mu} \right) - \frac{1}{2} q^2 \nabla^2 H_{\mu}  \right] dA~.
\label{eq:3.19}
\end{eqnarray}
This integral vanishes for the specific choice  $w = q$  since then we can combine
\begin{equation}
\frac{dZ}{dt}	=  \int \nabla \cdot \left( \frac{1}{2} q^2 \nabla H_{\mu} \right) dA = 0~.
\label{eq:3.20}
\end{equation}
Thus the self-adjoint operator 
\begin{equation}
\mathcal{A}_q (.) = \nabla \cdot (q \nabla (.))~,
\label{eq:3.20a}
\end{equation}
can be used to conserve total compressible kinetic and potential energy, as well as the potential enstrophy $Z$. 
%To complete the CLs we have to add gravity waves since we have assumed (\ref{eq:3.12})  for the conservation of $Z$. Thus the full set of equations for this process is
%\begin{eqnarray}
%\frac{\partial \zeta}{\partial t} = \mathcal{A} H_{\mu} ~, \\
%\frac{\partial \mu}{\partial t}  =  - \mathcal{A}  H_{\zeta} - \nabla^2 H_h~,\\
%\frac{\partial h}{\partial t}  =  \nabla^2 H_{\mu}~.
%\label{eq:3.21}
%\end{eqnarray}
Note that $\mathcal{A}$ is a functional of $q = Z_{\zeta}$.
This nonlinear process involves all variables: vorticity, divergence and height. The interaction between vortical and divergent kinetic energies is mediated by the potential vorticity $q$. This interaction is relevant for important processes in geophysical fluid dynamics like the emission of gravity waves by unbalanced flows or the equilibration towards balanced flow. Due to the different scales involved it is responsible for the energy transfer across scales.

%====================================
\subsection{Noncanonical Hamiltonian Form of the Equations}
%====================================

The combination of the terms identified above, i.e. of (\ref{eq:3.6}), (\ref{eq:3.8}), (\ref{eq:3.11}), (\ref{eq:3.12}), (\ref{eq:3.17}) and (\ref{eq:3.18}), with (\ref{eq:3.20a}), is the set of evolution equations for vorticity, divergence and height 
\begin{eqnarray}
\frac{\partial \zeta}{\partial t} = J(q,\chi) - \nabla \cdot (q \nabla \gamma) ~, \label{eq:3.22a} \\
\frac{\partial \mu}{\partial t}  =  J(q, \gamma) + \nabla \cdot (q \nabla \chi) - \nabla^2 \Phi~,\label{eq:3.22b} \\
\frac{\partial h}{\partial t}  =  - \nabla^2 \gamma~. \label{eq:3.22c}
\end{eqnarray}
The full system identifies the noncanonical Hamiltonian form of the equations and can be brought in a matrix notation with an anti-symmetric operator ${\boldsymbol A}$
\begin{equation}
\frac{\partial }{\partial t} 
 \left( \begin{array}{c}
  \zeta  \\
  \mu   \\
  h 
  \end{array}
  \right)
  = 
  {\boldsymbol A}
  \left( 
  \begin{array}{c}
  H_{\zeta}  \\
  H_{\mu}   \\
  H_h 
 \end{array}
 \right) ~,
\label{eq:3.23}
\end{equation}
where 
\begin{equation}
{\boldsymbol A}=
  \left(
 \begin{array}{ccc}
  -J_q & \mathcal{A}_q & 0  \\
  -\mathcal{A}_q & -J_q & -\nabla^2   \\
  0 & \nabla^2 & 0 
  \end{array}
  \right)
~,
\label{eq:3.23a}
\end{equation}
is the noncanonical Poisson operator and where $J_q f = J(q, f)$ is an antisymmetric operator for  a function $f$.
%Note that the $q$-dependent anti-symmetric operator $J_q f = J(q, f)$  for  a function $f$  appears on the diagonal of $J$. In (\ref{eq:3.23a}), $\mathcal{A}_q = \nabla \cdot (q \nabla(.))$  is self-adjoint and  $\nabla^2$  the symmetric Laplace operator.

%====================================
\section{Nambu and Poisson brackets}
\label{sec:3}
%====================================

The method presented in the previous sections leads directly to Hamiltonian and Nambu representations of the dynamics for both the vortical and divergent parts of the flow.

%====================================
\subsection{Vorticity Equation}
%====================================

The advection of vorticity by the vortical flow is
\begin{equation}
\frac{\partial \zeta}{\partial t}  = \{ \zeta , Z, H \}_{\zeta}~.
\label{eq:4.1}
\end{equation}
We can thus construct the Nambu bracket
\begin{equation}
\{F, Z, H \}_{\zeta} = \int F_{\zeta} J(H_{\zeta}, Z_{\zeta}) dA ~,
\label{eq:4.2}
\end{equation}
for the functional $F = F[\zeta]$. In (\ref{eq:4.2}), the subscript will be used to differentiate the bracket from the others representing the different processes.  The bracket (\ref{eq:4.2}) represents the 2D Euler equation. The bracket is anti-symmetric in all arguments, as it can be seen from  $\{F, Z, H \}_{\zeta} = - \{F, H, Z \}_{\zeta}$ due to the properties of the Jacobian, and is therefore cyclic, with $\{F, Z, H \}_{\zeta} = \{Z, H, F \}_{\zeta}$ and its permutations. The bracket (\ref{eq:4.2}) has quite a long history both in fluid dynamics and magneto hydrodynamics (MHD), see \cite{morrison1984bracket,bialynicki1991quantum,NevirBlender1993} and, more recently, \cite{andreussi2010mhd,bloch2013gradient,BlenderBadin15,blender2017viscous}.

The advection of the vorticity by the divergent flow yields a contribution to the divergence equation
\begin{equation}
\frac{\partial \mu}{\partial t}  = \{\mu, Z, H \}_{\mu}~,
\label{eq:4.3}
\end{equation}
where we have defined the bracket for $F = F[\mu]$ as
\begin{equation}
\{F, Z, H \}_{\mu}  =  \int F_{\mu} J(Z_{\zeta}, H_{\mu}) dA~.  
\label{eq:4.4}
\end{equation}
Equation (\ref{eq:4.3}) corresponds to the first term on the r.h.s. of (\ref{eq:3.22b}).
To construct a cyclic bracket we add the integral with cyclic arguments $(ZHF)$ and $(HZF)$ \cite{Salmon2007}, so that
\begin{equation}
\{ F, Z, H \}_{\mu}  = \int F_{\mu} J(Z_{\zeta}, H_{\mu}) dA + cyc(F,Z,H)~.
\label{eq:4.5}
\end{equation}
The two brackets (\ref{eq:4.4}) and (\ref{eq:4.5}) can be combined in a single Nambu bracket for $F = F[\zeta , \mu]$
\begin{eqnarray}
\{ F, Z, H \}_{\zeta \mu} =  \int F_{\zeta} J(Z_{\zeta}, H_{\zeta}) dA +  \int F_{\mu} J(Z_{\zeta}, H_{\mu}) dA \nonumber \\
= \{F, Z, H \}_{\zeta} + \{F, Z, H \}_{\mu}~.  
\label{eq:4.6}
\end{eqnarray}
The resulting bracket is based on the conservation of kinetic energy and potential enstrophy.

%====================================
\subsection{Gravity Waves}
%====================================

Gravity waves dynamics is described by the Poisson-bracket
\begin{eqnarray}
\frac{\partial \mu}{\partial t} =  \{ \mu, H \}_{\mu h}~,  \label{eq:4.7a} \\
\frac{\partial h}{\partial t} =  \{ h, H \}_{\mu h}~.  \label{eq:4.7b}
\end{eqnarray}
Both equations are combined for a functional $F[\mu,h]$ as
\begin{equation}
\{ F, H \}_{\mu h}  =  - \int F_{\mu} \nabla^2 H_h dA + \int F_h \nabla^2 H_{\mu} dA~.
\label{eq:4.8}
\end{equation}
The bracket is anti-symmetric, $\{F, H\}_{\mu h}   = -  \{ H, F \}_{\mu h}$.  
%Since the bracket possesses no Casimir, it is canonical. 
%Conservation of total energy is sufficient, potential enstrophy is not irrelevant. 
Note that no linearization is involved in the construction of (\ref{eq:4.8}).

%====================================
\subsection{Vorticity-Potential Flow Interaction}
%====================================

The interaction between vortical and potential flow can be described in Nambu form making use of the two equations
\begin{eqnarray}
\frac{\partial \zeta}{\partial t} = \nabla \cdot ( Z_{\zeta} \nabla H_{\mu})~, \label{eq:4.9a} \\
\frac{\partial \mu}{\partial t} = - \nabla \cdot ( Z_{\zeta} \nabla H_{\zeta})~. \label{eq:4.9b}
\end{eqnarray}

To determine the dynamics of a functional $F[\zeta]$, we employ (\ref{eq:4.9a}), to obtain
\begin{eqnarray}
\frac{\partial F}{\partial t}  = \{F, Z, H \}' = \int F_{\zeta} \nabla \cdot ( Z_{\zeta} \nabla H_{\mu}) dA \nonumber \\
=  - \int Z_{\zeta}  \left( \nabla F_{\zeta} \cdot \nabla H_{\mu} \right) dA~.
\label{eq:4.10}
\end{eqnarray}
In the last equality of (\ref{eq:4.10}) we have used the relationship  $\int \nabla \cdot (g q \nabla f) dA  =  0$  for arbitrary functions $g,~ q$ and $f$.

For a functional $F[\mu]$, we use instead (\ref{eq:4.9b}), which yields
\begin{eqnarray}
\frac{\partial F}{\partial t}  = \{ F, Z, H \}'' = - \int F_{\mu} \nabla \cdot ( Z_{\zeta} \nabla H_{\zeta} ) dA \nonumber \\
= \int Z_{\zeta} \left( \nabla F_{\mu} \cdot \nabla H_{\zeta} \right) dA~.
\label{eq:4.11}
\end{eqnarray}

The full bracket for arbitrary $F = F[\zeta , \mu]$ is thus written as
\begin{eqnarray}
\{ F, Z, H \}_{ \mu \zeta \zeta} =  \{ F, Z, H \}' + \{ F, Z, H \}''  \nonumber \\
= \int F_{\zeta} \nabla \cdot (Z_{\zeta} \nabla H_{\mu}) dA - \int F_{\mu} \nabla \cdot (Z_{\zeta} \nabla H_{\zeta}) dA \nonumber \\
=  - \int Z_{\zeta} \nabla F_{\zeta} \cdot \nabla H_{\mu}) dA  +  \int Z_{\zeta} \nabla F_{\mu} \cdot \nabla H_{\zeta}) dA~.  
\label{eq:4.12}
\end{eqnarray}
This bracket differs from Salmon's bracket \cite{Salmon2007} (see his equation (2.15)) 
%in two ways: first, our bracket does not include a singularity in $q$; secondly, 
since our bracket does not depend on a coordinate system. 
%The reasons behind these differences is that in (\ref{eq:3.20a}) we have replaced $q=Z_{\zeta}$, while Salmon's equation (2.15) relies on $q=-Z_h / q$.

The dynamics is thus determined by the conservation of both kinetic energy and potential vorticity. 
The equations of motion can be written in terms of the brackets introduced, as
\begin{eqnarray}
\frac{\partial \zeta}{\partial t}  = \{\zeta, Z, H\}_{\zeta \mu} + \{\zeta, Z, H \}_{\mu \zeta \zeta}~, \label{eq:4.13a} \\
\frac{\partial \mu}{\partial t}  = \{\mu, Z, H\}_{\zeta \mu} + \{\mu, Z, H\}_{\mu \zeta \zeta}  + \{\mu, H\}_{\mu h}~, \label{eq:4.13b} \\
\frac{\partial h}{\partial t}  =  \{h, H\}_{\mu h}~.  \label{eq:4.13c}
\end{eqnarray}
Equations (\ref{eq:4.13a})-(\ref{eq:4.13c}) consist of two Nambu-brackets, with two conserved integrals $Z$ and $H$, and one Poisson-bracket, with the Hamiltonian $H$ only. 
%Notice that the brackets can be combined independently. 

%====================================
\section{Summary}
\label{sec:4}
%====================================

The shallow water equations have been re-derived making use of geometric concepts, following the construction method previously applied by \cite{BlenderBadin15} to the generalized 2D Euler and to the Rayleigh-B{\'e}nard equations. The main advantage of the method presented here is to simplify the intuitive approach which is often necessary in the construction of dynamical equations consistent with conservation laws.

The derivation of the dynamics is based on the idea to use CLs as constraints for the trajectories in state space. The vorticity dynamics in the shallow water approximation is thus constructed using exterior differential forms.  In agreement with the 2D Euler equations \cite{BlenderBadin15}, the energy and potential enstrophy appear symmetric in a Nambu bracket. The main extension compared to \cite{BlenderBadin15} is to identify the energy conserving process using self-adjoint operators. Individual mechanisms are selected by the corresponding energy transformations. 
Potential enstrophy conservation is used as a constraint when vortical flows are involved. The dynamics is determined by the Hamiltonian and the functional dependencies on the dynamical variables.

%A main advantage of the method presented here is to simplify the intuitive approach which is often necessary in the construction of dynamical equations consistent with CLs. 

Our formulation for the Nambu bracket differs from the bracket found by \cite{Salmon2007} 
\begin{eqnarray}
\{ F,H,Z_n \}_{\mu \mu \zeta} = \frac{1}{3} \int  \left[ J (F_\mu,  H_\mu)  Z_\zeta + cyc (F, H,  Z) \right] dA~, 
\label{eq:salmon1}
\end{eqnarray}
\begin{eqnarray}
\{ F,H,Z \}_{\zeta \zeta \zeta} = \frac{1}{3} \int \left[ J (F_\zeta, H_\zeta)  Z_\zeta + cyc (F, H, Z) \right] dA~, 
\label{eq:salmon2}
\end{eqnarray}
\begin{eqnarray}
\{ F,H,Z \}_{\zeta \mu h} = -  \int  \left( \frac{\partial_x F_\mu \, \partial_x H_\zeta - \partial_x H_\mu \, \partial_x F_\zeta}{\partial_x q} \partial_x Z_h  \right.
      \nonumber \\
+ \left. \frac{\partial_y F_\mu \partial_y H_\zeta - \partial_y  H_\mu  \partial_y F_\zeta}{\partial_y q} \partial_y Z_h  \right) dA + cyc (F, H, Z)~,
\label{eq:salmon3}
\end{eqnarray}
as it does not
%in the way in which Salmon's bracket are singular as they include $q$ in the denominator, and they 
depend on the coordinate system.
It should be noted that in (\ref{eq:salmon1})-(\ref{eq:salmon3}) the sum of the three brackets comprises the complete Nambu bracket for
the shallow water equations. 

%%(see his equation (2.15)) 
%The comparison between the brackets found in this study and (\ref{eq:salmon1}-(\ref{eq:salmon3} show that the as it does not
%%in the way in which Salmon's bracket are singular as they include $q$ in the denominator, and they 
%depend on the coordinate system. 
%\cite{Salmon2007} observed that the numerical solution of the equations obtained from the discretization of the bracket do not show singularities. In our case, the problem is eliminated by the non-singular form of the brackets themselves.
Our approach demonstrates thus that two CLs and three dynamical variables are sufficient to setup the shallow water model. 
%One might even conclude that a hydrodynamic system like this model is completely determined by its CLs. Geometrically this is obvious since the dynamics in phase space has to take place on the intersections of the conserved integrals. 
The equations are determined up to a constant factor.
%not unique since the evolution along the trajectories is not fully determined. In our derivation this caused no severe problem. 

Note that we did not identify constitutive CLs which yield modular equations and possible approximations. Such conserved integrals, which have been called constitutive CLs \cite{NevirSommer2009}, determine the dynamics but are not conserved in the full system. It is possible that the identification of appropriate conserved integrals and subsystems allows a clearer separation of processes. 

%%%%%%%%%%%%%%%%%%%%%%%%%%%%%%%%%%%%%%%%%%
\vspace{6pt} 

%%%%%%%%%%%%%%%%%%%%%%%%%%%%%%%%%%%%%%%%%%
%% optional
%\supplementary{The following are available online at www.mdpi.com/link, Figure S1: title, Table S1: title, Video S1: title.}

%%%%%%%%%%%%%%%%%%%%%%%%%%%%%%%%%%%%%%%%%%
%\acknowledgments{The authors would like to acknowledge three anonymous referees for comments that helped to improve the manuscript. This research was partially funded by the research grant DFG TRR181.}
%
%%%%%%%%%%%%%%%%%%%%%%%%%%%%%%%%%%%%%%%%%%%
%\authorcontributions{Both authors contributed equally to the development of the ideas here presented.}
%
%%%%%%%%%%%%%%%%%%%%%%%%%%%%%%%%%%%%%%%%%%%
%\conflictsofinterest{The authors declare no conflict of interest.} 

%%%%%%%%%%%%%%%%%%%%%%%%%%%%%%%%%%%%%%%%%%
%% optional
%\abbreviations{The following abbreviations are used in this manuscript:\\
%
%\noindent 
%\begin{tabular}{@{}ll}
%MDPI & Multidisciplinary Digital Publishing Institute\\
%DOAJ & Directory of open access journals\\
%TLA & Three letter acronym\\
%LD & linear dichroism
%\end{tabular}}

%%%%%%%%%%%%%%%%%%%%%%%%%%%%%%%%%%%%%%%%%%
%% optional
%\appendixtitles{no} %Leave argument "no" if all appendix headings stay EMPTY (then no dot is printed after "Appendix A"). If the appendix sections contain a heading then change the argument to "yes".
%\appendixsections{one} %Leave argument "multiple" if there are multiple sections. Then a counter is printed ("Appendix A"). If there is only one appendix section then change the argument to "one" and no counter is printed ("Appendix").
\appendix
\section{Nambu form of a finite dimensional fast-slow shallow water system}
\label{sec:5}
%====================================

Lorenz \cite{lorenz86} introduced the following set of five ordinary differential equations 
\begin{equation}
\frac{d x_1}{d t} = -x_2 x_3 + b x_2 x_5~,
\label{eq:A1}
\end{equation}
\begin{equation}
\frac{d x_2}{d t} = x_1 x_3 - b x_1 x_5~, 
\label{eq:A1b}
\end{equation}
\begin{equation}
\frac{d x_3}{d t} = -x_1 x_2 ~, 
\label{eq:A1c}
\end{equation}
\begin{equation}
\frac{d x_4}{d t} = -\frac{x_5}{\epsilon}~, 
\label{eq:A1d}
\end{equation}
\begin{equation}
\frac{d x_5}{d t} = -\frac{x_4}{\epsilon} + b x_1 x_2~. 
\label{eq:A1e}
\end{equation}				 
System (\ref{eq:A1})-(\ref{eq:A1e}) can be derived from a Fourier expansion of the shallow water equations and describes the interaction between a vorticity triad  $(x_1 , x_2, x_3)$, also called "slow" variables, with a gravity wave mode with divergence and geostrophic imbalance with coefficients, respectively, $x_4$ and $x_5$, also called "fast" variables. In (\ref{eq:A1})-(\ref{eq:A1e}), $b$ is the rotational Froude number, responsible for the coupling between the slow and fast modes, and  $\epsilon$ is a frequency separation parameter, proportional to the Rossby number, introduced by \cite{BokhoveShepherd96}. The term "separation parameter" becomes clear in the case $\epsilon \ll 1$, which corresponds to a formal separation of time scales between the slow and fast modes. A lot of work has been done on the system (\ref{eq:A1})-(\ref{eq:A1e}), in particular to try to understand if it allows for a formal separation of the manifolds over which the slow and fast dynamics take place \cite{lorenz86,LorenzKrishnamurthy87,Lorenz92,vallis1992mechanisms,Camassa95,BokhoveShepherd96,Vanneste04}; see \cite{Vanneste13} for a review and further references. It is curious to note that the system itself is analogous to a swinging spring, i.e. a pendulum in which the mass is attached to a spring \cite{Lynch02}. In this Appendix we will introduce the Nambu form for system (\ref{eq:A1})-(\ref{eq:A1e}). Following \cite{BlenderBadin15}, the dynamics of a finite dimensional system characterized by a Hamiltonian $H$ and a Casimir $Z$ can be written in Nambu form as
\begin{equation}
\frac{d}{dt}\vec{x}=\nabla Z \times \nabla H~,
\label{eq:A0}
\end{equation} 				 
so that, an arbitrary state space function $F(\vec{x})$ is given by the canonical braket
\begin{equation}
\frac{\partial F}{ \partial t}=\nabla F \cdot \nabla Z \times \nabla H \equiv \{ F,Z,H \}~.
\label{eq:A0b}
\end{equation} 
To put (\ref{eq:A1})-(\ref{eq:A1e}) in form (\ref{eq:A0}), start by noticing that the system conserves the total Hamiltonian
\begin{equation}
H=\frac{1}{2}(x^2_1+2 x^2_2 + x^2_3 + x^2_4 +x^2_5)~,
\label{eq:A2}
\end{equation} 				 
and the total enstrophy
\begin{equation}
Z=\frac{1}{2}(x^2_2 + x^2_3 + x^2_4 +x^2_5)~.
\label{eq:A3}
\end{equation}  				 
The uncoupled system, obtained by setting  $b=0$, has Hamiltonian and enstrophy, for the slow component, respectively
\begin{equation}
H_{123}=\frac{1}{2}(x^2_1+2 x^2_2 + x^2_3)~,
\label{eq:A4}
\end{equation} 			  
and 
\begin{equation}
Z_{123}=w H_{123} - \frac{1}{2}(x^2_1 + x^2_2)~.
\label{eq:A5}
\end{equation} 				 
In this case, the uncoupled case reduces instead to a harmonic oscillator with frequency $1/\epsilon$ \cite{BokhoveShepherd96}. The coupling between the vorticity and ageostrophic mode introduces instead a Hamiltonian and an enstrophy with the same forms of (\ref{eq:A4}) and (\ref{eq:A5}), but with the dependence on  $x_3$ replaced by $x_5$, so that    
\begin{equation}
H_{125}=\frac{1}{2}(x^2_1+2 x^2_2 + x^2_5),~Z_{125}=w H_{125} - \frac{1}{2}(x^2_1 + x^2_2)~.
\label{eq:A6}
\end{equation}		  
where $w \in \mathbb{R}$  is arbitrary.
With these considerations, the dynamics of (\ref{eq:A1})-(\ref{eq:A1e}) can be written as
\begin{equation}
\frac{d}{dt} \vec{x} = \nabla_{123} Z_{123} \times \nabla_{123} H_{123} - b \nabla_{125} Z_{125} \times \nabla_{125} H_{125} + \frac{1}{\epsilon} [\vec{x} , H_{45}]_{45}~,
\label{eq:A7}
\end{equation} 		 
i.e., it can be written as the sum of two Nambu brackets and one Poisson bracket, that is the same structure that was found for the infinite dimensional system (\ref{eq:4.13a})-(\ref{eq:4.13c}). In (\ref{eq:A7}) we have set  $\nabla_{ijk}=(\partial_i, \partial_j, \partial_k)$ and $[F,G]_{ij}=(\partial_i F \partial_j G - \partial_j F \partial_i G)$  for arbitrary phase space functions $F(\vec{x})$  and $G(\vec{x})$. (\ref{eq:A7}) can thus be generalized as
\begin{equation}
\frac{d}{dt} F(\vec{x}) = \{ F, Z_{123}, H_{123} \}_{123} - b \{ F, Z_{125}, H_{125} \}_{125} + \frac{1}{\epsilon} [F , H_{45}]_{45}~,
\label{eq:A8}
\end{equation} 		  
where 
\begin{equation}
\{ F, Z_{12i}, H_{12i} \}_{12i} = \nabla_{12i} F \cdot \nabla_{12i} Z_{12i} \times \nabla_{12i} H_{12i} ~,~i=3,5~,
\label{eq:A9}
\end{equation} 		 
are the Nambu brackets of the system.
%==================
%\ack{This research was partially funded by the research grant DFG TRR181.}

\clearpage

\bibliographystyle{unsrt}
%\bibliography{references}

%\section*{References}
%\begin{thebibliography}{10}
%\bibitem{book1} Goosens M, Rahtz S and Mittelbach F 1997 {\it The \LaTeX\ Graphics Companion\/} 
%(Reading, MA: Addison-Wesley)
%\bibitem{eps} Reckdahl K 1997 {\it Using Imported Graphics in \LaTeX\ } (search CTAN for the file `epslatex.pdf')
%\end{thebibliography}

\end{document}